\documentclass[conference]{IEEEtran}
\IEEEoverridecommandlockouts

\usepackage{cite}
\usepackage{makeidx}
\usepackage{url}
\usepackage{graphicx}
\usepackage{enumitem}
\usepackage{amsmath}
\usepackage{amsfonts}
\usepackage{amssymb}
\usepackage{amsthm}
\usepackage{algorithmic}
\usepackage{booktabs}
\usepackage{tabularx}
\usepackage{makecell}
\usepackage{float}
\usepackage{mathtools}
\usepackage{caption}
\usepackage{subcaption}
\usepackage{bbm}
\usepackage[most]{tcolorbox}

\theoremstyle{definition}

\newcommand{\refig}[1]{Fig.~\ref{#1}}
\newcommand{\refsec}[1]{Sec.~\ref{#1}}

\usepackage[ruled,vlined,linesnumbered]{algorithm2e}
\SetAlgorithmName{Algorithm}{Algorithm}{List of Algorithms}

\usepackage{makecell}
\usepackage{multirow}

\begin{document}

\title{MambaCSP: Hybrid-Attention State Space Models for Hardware-Efficient Channel State Prediction
\thanks{
This work acknowledges support from the German BMFTR through the 6G-life (16KISK002), 6GEM+ (16KIS2409K), GEM-X (16KISS004K), VICTOR6G (16KIS2547), QD-CamNetz (16KISQ077), QuaPhySI (16KIS1598K), QUIET (16KISQ093), and AISAC (16KIS2462) projects, as well as from the German DFG through the CeTI project (Germany’s Cluster of Excellence -- EXC 2050/2, ID 390696704).
}
}

\author{%
    \IEEEauthorblockN{%
        Aladin~Djuhera\IEEEauthorrefmark{1},
        Haris~Gacanin\IEEEauthorrefmark{2}, 
        Holger~Boche\IEEEauthorrefmark{1}
    }
    \IEEEauthorblockA{%
        \IEEEauthorrefmark{1}Technical University Munich, Germany,\;
        \IEEEauthorrefmark{2}RWTH Aachen University, Germany\\
        Emails: \{aladin.djuhera, boche\}@tum.de, harisg@dsp.rwth-aachen.de
    }
}

\maketitle


\begin{abstract}
    Recent works have demonstrated that attention-based transformer and large language model (LLM) architectures can achieve strong channel state prediction (CSP) performance by capturing long-range temporal dependencies across channel state information (CSI) sequences.
    However, these models suffer from quadratic scaling in sequence length, leading to substantial computational cost, memory consumption, and inference latency, which limits their applicability in real-time and resource-constrained wireless deployments.
    In this paper, we investigate whether selective state space models (SSMs) can serve as a hardware-efficient alternative for CSI prediction.
    We propose MambaCSP, a hybrid-attention SSM architecture that replaces LLM-based prediction backbones with a linear-time Mamba model.
    To overcome the local-only dependencies of pure SSMs, we introduce lightweight patch-mixer attention layers that periodically inject cross-token attentions, helping with long-context CSI prediction. 
    Extensive MISO-OFDM simulations show that MambaCSP improves prediction accuracy over LLM-based approaches by 9--12\%, while delivering up to 3.0x higher throughput, 2.6x lower VRAM usage, and 2.9x faster inference.
    Our results demonstrate that hybrid state space architectures provide a promising direction for scalable and hardware-efficient AI-native CSI prediction in future wireless networks.
\end{abstract}



\section{Introduction and Motivation}
\label{sec:introduction}

Large language models (LLMs) have gained strong capabilities in coding, math, and complex reasoning, demonstrating their transformative potential across diverse domains.
In future 6G networks, they can serve as foundational building blocks for embedding agentic capabilities across communication, connectivity, and artificial intelligence (AI) layers, thereby replacing or augmenting existing workflows \cite{jiang2026large, djuhera2026monolithicwirelessworldmodel,djuhera2025jointpartitioningplacementfoundation}.

However, not all communication layers can readily benefit from such AI-native integration.
Wireless signal processing systems operate under strict real-time constraints, which have traditionally limited the adoption of AI due to high computational and latency requirements \cite{guo2026large}.
LLMs, in particular, contain billions of parameters and require substantial compute, making them unsuitable for real-time deployment on resource-constrained devices and in latency-sensitive applications where time to first token (TTFT) is critical.
Overcoming these constraints is essential for enabling practical AI-native integration.

Among such latency-sensitive applications is channel state prediction (CSP), which forecasts future channel state information (CSI) from historical observations, thereby reducing overall pilot overhead. 
However, acquiring CSI via conventional channel estimation methods is challenging, particularly in high-mobility scenarios where shortened channel coherence times significantly increase the estimation overhead. 
Recent works \cite{liu2024llm4cp, fan2025csi, cui2025exploring} have explored LLM-based architectures for CSP, leveraging attention mechanisms to effectively capture long-range, non-local dependencies across time and frequency. 
Despite their strong performance, transformer-based models exhibit quadratic scaling in sequence length, resulting in substantial computational and memory overhead, as well as increased inference latency due to key-value (KV) caching.
This makes them impractical for long-context CSI prediction and real-time deployment in wireless systems.

Selective state space models (SSMs), such as Mamba \cite{gu2024mamba}, offer a hardware-efficient alternative to LLMs by modeling sequences without KV caching, resulting in linear complexity.
In particular, state transitions are input-dependent and include a continuous-time inductive bias, making SSMs better suited for modeling wireless channel variations.
However, pure SSMs process tokens recursively, which can limit their ability to capture non-local dependencies in complex CSI sequences.

In this paper, we address these limitations and propose \textbf{MambaCSP}, a hybrid-attention state space architecture for hardware-efficient CSP.
Our approach employs a Mamba-based backbone with lightweight patch-mixer attention layers that periodically inject cross-token attention, enabling modeling of long-range dependencies across CSI sequences while preserving the efficiency of SSMs.
To this end, we adapt and generalize LLM-based CSP pipelines into a unified framework that enables efficient processing of CSI sequences across both model architectures.
Our key findings are:
\begin{itemize}

    \item \textbf{Prediction Performance}: MambaCSP consistently outperforms LLM-based and other neural baselines across diverse mobility scenarios in both TDD and FDD settings.

    \item \textbf{Hardware Efficiency}: MambaCSP achieves up to 3.0$\times$ higher throughput, 2.6$\times$ lower memory usage, and 2.9$\times$ faster inference compared to LLM-based prediction.

    \item \textbf{Hybrid-Attention Design}: MambaCSP's patch-mixer attention is particularly beneficial in FDD scenarios, where non-local, cross-frequency dependencies require explicit token interactions beyond recursive state-space modeling.  
    
\end{itemize}

This paper is organized as follows.
Section II presents the system model and formalizes the problem statement. 
Section III introduces the CSI prediction pipeline and MambaCSP. 
Section IV describes the experiment setup. 
Section V discusses the simulation results, and Section VI concludes the paper.
\section{System Model and CSI Prediction Task}
\label{sec:system_model}

In this section, we formally define the wireless system model and the corresponding channel state prediction task.

\subsection{Wireless System Model}

We consider a single-cell MISO-OFDM link, where a base station (BS) equipped with a uniform planar array (UPA) with $N_t$ antennas serves $U$ single-antenna user equipments (UEs).
We follow a 5G NR FR1 OFDM numerology with carrier frequency $f_c = 2.4$~GHz, subcarrier spacing $\Delta f = 15$ kHz, and $N_{\mathrm{sc}} = 12$ subcarriers per resource block (RB), such that each RB has a bandwidth of $B_{\mathrm{RB}} = N_{\mathrm{sc}} \Delta f = 180$ kHz.
We consider two contiguous frequency bands of equal bandwidth for uplink (UL) and downlink (DL), each consisting of $K_{\mathrm{UL}} = K_{\mathrm{DL}} = 48$ RBs, resulting in a per-band bandwidth of $8.64$ MHz.
Time is organized in slots of duration $T_{\mathrm{slot}} = 1$ ms and users move with velocities in the range of $v \in [10,100]$ km/h.

For each UE, we consider a sequence of $N_{\mathrm{slot}} = P + L$ consecutive slots, where the first $P$ slots form a \emph{history window} and the remaining $L$ slots form a \emph{prediction window}.
For the UL band, let $k \in \{1,\dots,K_{\mathrm{UL}}\}$ denote the RB index and $s \in \{1,\dots,N_{\mathrm{slot}}\}$ the slot index.
The corresponding narrowband MISO channel vector on RB $k$ and slot $s$ is denoted as
\begin{equation}
    \mathbf{h}^{\mathrm{UL}}_{k,s} \in \mathbb{C}^{N_t} \ .
\end{equation}
Similarly, for the DL band, we define $\mathbf{h}^{\mathrm{DL}}_{k,s} \in \mathbb{C}^{N_t}$.

We model the wireless channel as frequency-selective and time-varying, incorporating multipath propagation and Doppler effects induced by user mobility.
The corresponding parametric representation of the UL channel is thus given by
\begin{equation}
    \mathbf{h}^{\mathrm{UL}}_{k,s} = \sum_{\ell=1}^{L_p} \alpha_{\ell} 
    \mathbf{a}_{\mathrm{BS}}(\theta_\ell,\phi_\ell)
    e^{-j2\pi f_k \tau_\ell} e^{j2\pi f_{D,\ell} t_s} \ ,
\end{equation}
where $\alpha_\ell$, $\tau_\ell$, and $(\theta_\ell,\phi_\ell)$ denote the complex gain, delay, and angles of departure of path $\ell$, $\mathbf{a}_{\mathrm{BS}}(\cdot)$ is the BS array response, and $f_{D,\ell}$ is the Doppler shift.
We consider both time-division duplex (TDD) and frequency-division duplex (FDD) modes.

\subsection{CSI-RS and DMRS Configuration}

Channel estimation in 5G NR follows the \emph{Demodulation Reference Signal} (DMRS) pilot pattern, where pilots are transmitted on a sparse subset of RBs and symbols.
Formally, let $\mathcal{K}_{\mathrm{CSI}} \subseteq \{1,\dots,K_{\mathrm{UL}}\}$ denote the set of RBs carrying pilot signals and let $\mathcal{S}_{\mathrm{CSI}} \subseteq \{1,\dots,P\}$ denote the corresponding time indices.
The observed CSI is given by
\begin{equation}
    \tilde{\mathbf{h}}^{\mathrm{UL}}_{k,s} = \mathbf{h}^{\mathrm{UL}}_{k,s} + \mathbf{n}_{k,s} \ ,
    \quad (k,s) \in \mathcal{K}_{\mathrm{CSI}} \times \mathcal{S}_{\mathrm{CSI}} \ ,
\end{equation}
where $\mathbf{n}_{k,s} \sim \mathcal{CN}(0,\sigma_n^2 \mathbf{I})$ models estimation noise.
Note that different DMRS pilot patterns can be configured depending on the scenario, controlling the corresponding pilot density.

\subsection{Channel State Prediction Task}

Given the UL history window over $P$ slots, we aim to predict the future DL CSI over the next $L$ slots across all RBs.
To this end, let the UL history be defined as
\begin{equation}
    \mathcal{H}_{\mathrm{his}}^{\mathrm{UL}} = 
    \{\tilde{\mathbf{h}}^{\mathrm{UL}}_{k,s} \mid k \in \mathcal{K}_{\mathrm{CSI}}, s \in \mathcal{S}_{\mathrm{CSI}}\} \ 
\end{equation}
with corresponding DL prediction target
\begin{equation}
    \mathcal{H}_{\mathrm{pre}}^{\mathrm{DL}} = 
    \{\mathbf{h}^{\mathrm{DL}}_{k,s} \mid k \in \{1,\dots,K_{\mathrm{DL}}\} , s \in \{P+1,\dots,P+L\} \} \ ,
\end{equation}
This defines a sequence-to-sequence prediction of the form 
\begin{equation}
    \hat{\mathcal{H}}_{\mathrm{pre}}^{\mathrm{DL}} 
    = f_{\Theta}(\mathcal{H}_{\mathrm{his}}^{\mathrm{UL}}) \ ,
\end{equation}
where $f_{\Theta}$ denotes a parameterized model with learnable parameters $\Theta$, whose prediction performance is evaluated using the normalized mean square error (NMSE), i.e.,
\begin{equation}
    \mathrm{NMSE} =
    \mathbb{E} \left[
    \frac{\sum_{k,s} \|\hat{\mathbf{h}}^{\mathrm{DL}}_{k,s} - \mathbf{h}^{\mathrm{DL}}_{k,s}\|_2^2}
         {\sum_{k,s} \|\mathbf{h}^{\mathrm{DL}}_{k,s}\|_2^2}
    \right].
\end{equation}

In our work, we require $f_{\Theta}$ to not only accurately predict future DL CSI, but also to minimize the computational complexity and latency, enabling real-time deployment.
\section{MambaCSP: Hybrid-Attention State Space Models for CSI Prediction}
\label{sec:MambaCSP}

\begin{figure*}
    \centering
    \includegraphics[width=1\linewidth]{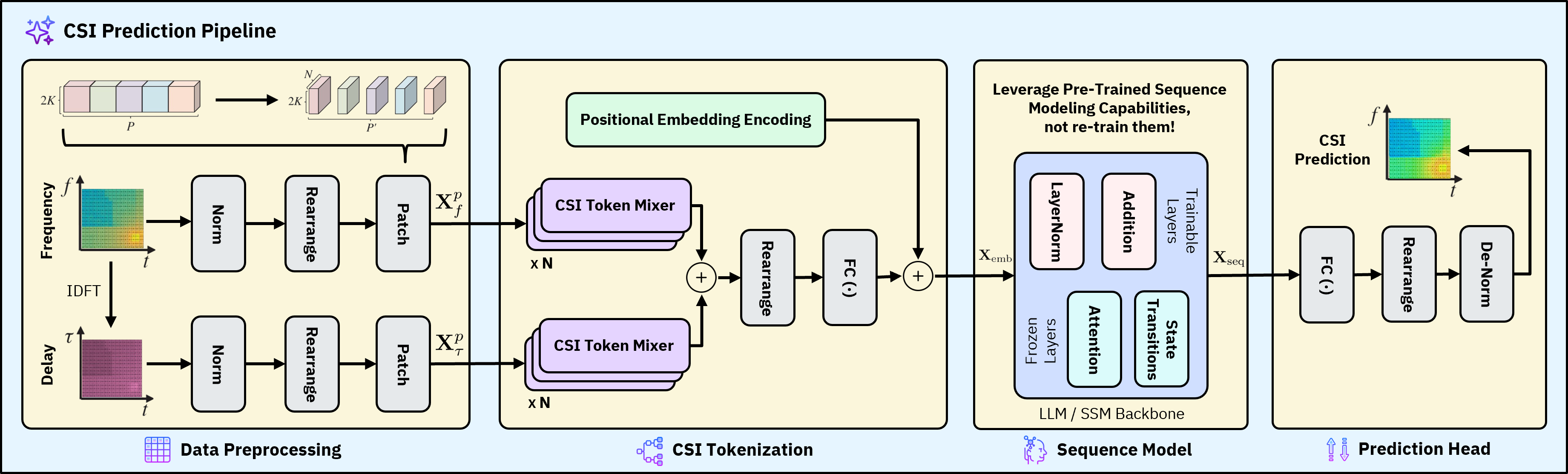}
    \caption{CSI prediction pipeline: Historical UL CSI is converted into frequency/delay components, normalized, rearranged, and partitioned into temporal patches. CSI token mixer blocks then produce token embeddings for the sequence model. A final prediction head maps the outputs back to the complex CSI domain.}
    \label{fig:CSI_pipeline}
\end{figure*}

In this section, we introduce a unified framework for CSI prediction that is compatible with both LLM and SSM sequence models.
We then present the proposed hybrid MambaCSP architecture, which combines the efficiency of SSMs with the expressivity of attention mechanisms.

\subsection{Unified CSI Prediction Framework}

Building on prior works \cite{liu2024llm4cp, fan2025csi}, we generalize attention-based CSP pipelines into a unified framework that supports arbitrary sequence modeling architectures, including SSMs (see \refig{fig:CSI_pipeline}).
The key components are outlined as follows.

\subsubsection{Data Preprocessing}

CSI data is complex-valued and needs to be converted into real-valued embeddings first.
To this end, we extract frequency- and delay-domain representations, normalize the data, and partition it into temporal patches.
We denote the historical UL CSI over $P$ slots and $K$ RBs as $\mathbf{H}_f \in \mathbb{C}^{K \times P}$ and its delay-domain representation as $\mathbf{H}_\tau = \mathbf{F}_K^{H} \mathbf{H}_f$, where $\mathbf{F}_K$ denotes the $K$-point DFT matrix.
We then separate real and imaginary components and obtain $\mathbf{X}_f, \mathbf{X}_\tau \in \mathbb{R}^{2 \times K \times P}$.
To ensure stable training across varying SNR conditions, we normalize each representation and reshape the tensors by merging feature dimensions, yielding $\tilde{\mathbf{X}}_f, \tilde{\mathbf{X}}_\tau \in \mathbb{R}^{2K \times P}$.
To further reduce sequence length and computational complexity, we partition the temporal dimension into non-overlapping patches of size $N$ (see \refig{fig:CSI_pipeline}a), resulting in
\begin{equation}
    \mathbf{X}_f^{p}, \mathbf{X}_\tau^{p} \in \mathbb{R}^{2K \times N \times P'},
\end{equation}
where $P' = \lceil P / N \rceil$ denotes the number of patches.

\subsubsection{CSI Tokenization}

\begin{figure}[t]
    \centering
    \includegraphics[width=0.98\linewidth]{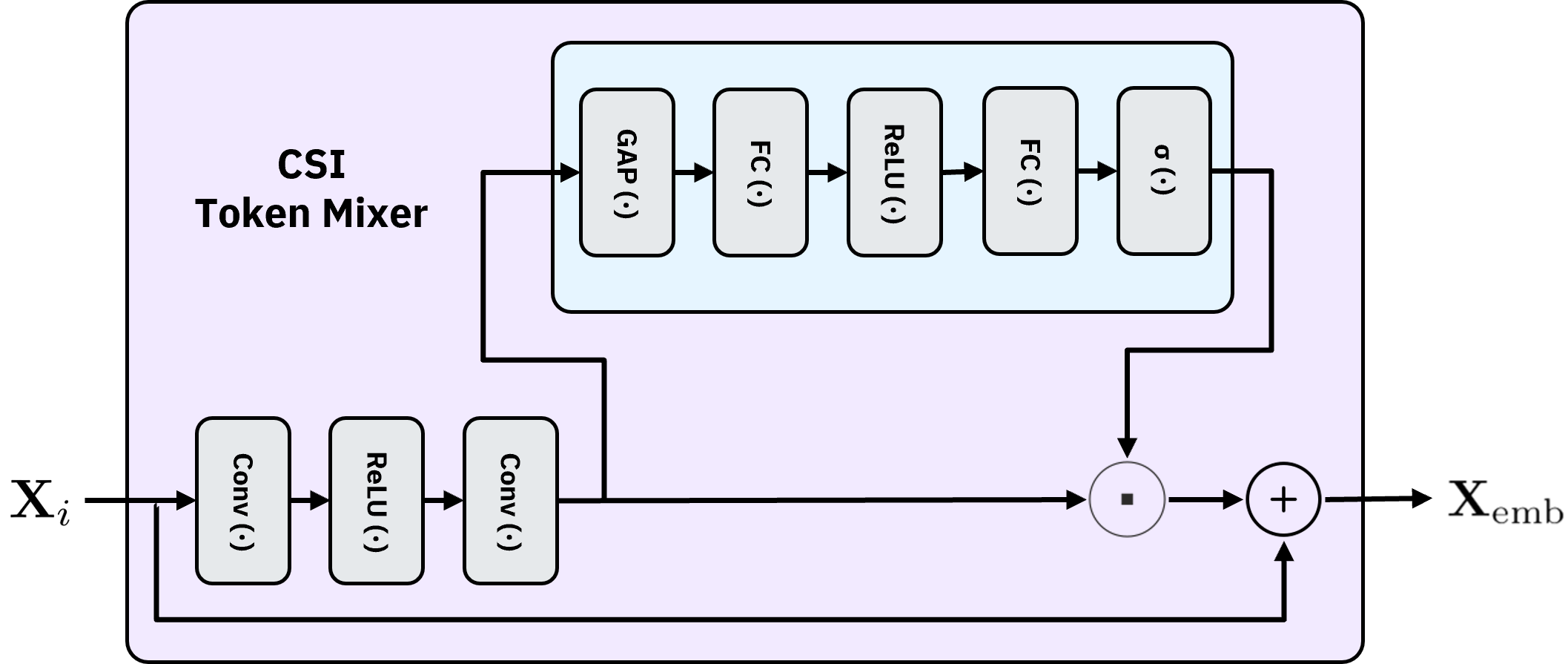}
    \caption{CSI token mixer module.}
    \label{fig:CSI_token_mixer}
    \vspace{-0.5cm}
\end{figure}

Preprocessed CSI patches need to be further tokenized into embeddings that are compatible with the input format of language models.
To this end, each patch is processed by a \emph{CSI Token Mixer} (see \refig{fig:CSI_token_mixer}), which adaptively reweights features to emphasize informative multipath components while attenuating less relevant or noisy ones.

Let $\mathbf{X}_i \in \mathbb{R}^{2K \times N \times P'}$ denote the input tensor of a token mixer block.
First, local feature extraction via convolutional $\mathrm{Conv}(\cdot)$ and $\mathrm{ReLU}(\cdot)$ layers is performed to capture joint temporal and frequency-domain dependencies, i.e.,
\begin{equation}
    \mathbf{X}_{\mathrm{feat}} = \mathrm{Conv}(\mathrm{ReLU}(\mathrm{Conv}(\mathbf{X}_i))) \ .
\end{equation}

To model the relative importance of each token, a channel-wise gating mechanism with sigmoid activation $\sigma(\cdot)$ is applied
\begin{equation}
    \mathbf{X}_{\mathrm{gate}} = \sigma(\mathrm{FC}(\mathrm{ReLU}(\mathrm{FC}(\mathrm{GAP}(\mathbf{X}_{\mathrm{feat}}))))) \ ,
\end{equation}
where $\mathrm{GAP}(\cdot)$ and $\mathrm{FC}(\cdot)$ denote global average pooling and fully connected layers.
The resulting weights $\mathbf{X}_{\mathrm{gate}} \in \mathbb{R}^{1 \times 1 \times P'}$ are then used to reweight each token, i.e.,
\begin{equation}
    \mathbf{X}_{\mathrm{sca}}[:,:,i] = \mathbf{X}_{\mathrm{gate}}[i] \cdot \mathbf{X}_{\mathrm{feat}}[:,:,i] \ ,
\end{equation}
before applying a residual connection, which preserves the original features and stabilizes gradient flow during training
\begin{equation}
    \mathbf{X}_o = \mathbf{X}_{\mathrm{sca}} + \mathbf{X}_i \ .
\end{equation}

Next, to jointly capture spectral correlations and multipath structure, we cascade multiple CSI Token Mixer blocks over the frequency/delay representations, and combine them as
\begin{equation}
    \mathbf{X}_{\mathrm{tok}} = \mathrm{TM}^{(N)}(\mathbf{X}_f^{p}) + \mathrm{TM}^{(N)}(\mathbf{X}_\tau^{p}) \ ,
\end{equation}
where $\mathrm{TM}^{(N)}(\cdot)$ denotes $N$ cascaded token mixer blocks.
In practice, we employ $N \in [2,4]$ to balance modeling capacity and computational efficiency.
The resulting tensor is rearranged into $\tilde{\mathbf{X}}_{\mathrm{tok}} \in \mathbb{R}^{2KN \times P'}$ and projected into the model embedding dimension, yielding $\bar{\mathbf{X}}_{\mathrm{tok}} \in \mathbb{R}^{F \times P'}$.

Furthermore, to preserve the temporal ordering, we follow \cite{liu2024llm4cp} and add sinusoidal positional encodings defined as
\begin{equation}
    \mathbf{X}_{\mathrm{PE}}(i,j) =
    \begin{cases}
        \sin\left(\frac{j}{10000^{i/F}}\right) \ , & i \ \text{even} \ , \\
        \cos\left(\frac{j}{10000^{(i-1)/F}}\right) \ , & i \ \text{odd} \ ,
    \end{cases}
\end{equation}
where $i$ indexes the feature dimension and $j$ the token position.
The final CSI token embeddings are then obtained as
\begin{equation}
    \mathbf{X}_{\mathrm{emb}} = \bar{\mathbf{X}}_{\mathrm{tok}} + \mathbf{X}_{\mathrm{PE}} \ .
\end{equation}

\subsubsection{Sequence Model Backbone}

To model spatio-temporal dependencies across CSI token sequences, we leverage the pre-trained backbone's capacity to attend to complex multi-scale interactions rather than its learned semantic world knowledge.
During training, we therefore keep the backbone's core layers frozen while only adapting lightweight task-specific components.
In LLMs, for example, this corresponds to freezing the multi-head attention and feed-forward blocks, while only fine-tuning LayerNorm and Addition layers \cite{liu2024llm4cp, fan2025csi}.
This significantly reduces the overall training overhead and yields contextualized token representations given by
\begin{equation}
    \mathbf{X}_{\mathrm{seq}} = f_{\Theta_{\mathrm{bb}}}(\mathbf{X}_{\mathrm{emb}}) \in \mathbb{R}^{F \times P'} \ ,
\end{equation}
where $f_{\Theta_{\mathrm{bb}}}(\cdot)$ denotes the sequence model backbone, and $\mathbf{X}_{\mathrm{seq}}$ captures the learned spatio-temporal dependencies.

\subsubsection{Prediction Head}

In the final step, $\mathbf{X}_{\mathrm{seq}}$ is mapped back to its original domain.
We first apply a linear projection to obtain $\hat{\mathbf{X}} \in \mathbb{R}^{2K \times L}$ and reshape the predicted tensor into real/imaginary components of the form $\hat{\mathbf{X}} \in \mathbb{R}^{2 \times K \times L}$.
We then de-normalize to recover the original CSI scale, i.e.,
\begin{equation}
    \hat{\mathbf{X}}_{\mathrm{de}} = \sigma \, \hat{\mathbf{X}} + \mu \ ,
\end{equation}
where $\mu$ and $\sigma$ denote the mean and standard deviation used during preprocessing.
Finally, the predicted CSI is obtained as
\begin{equation}
    \hat{\mathbf{H}}^{\mathrm{DL}} = 
    \hat{\mathbf{X}}_{\mathrm{de}}[1,:,:] 
    + j \, \hat{\mathbf{X}}_{\mathrm{de}}[2,:,:] \ .
\end{equation}

\subsection{Hybrid Mamba Architecture}

Mamba~\cite{gu2024mamba} extends classical SSMs by introducing input-dependent state transitions. 
Instead of fixed matrices $\mathbf{A}$ (state transition), $\mathbf{B}$ (input projection), and $\mathbf{C}$ (output projection), Mamba parameterizes these operators as functions of the input:
\begin{equation}
\mathbf{h}_t = \mathbf{A}(\mathbf{x}_t)\mathbf{h}_{t-1} + \mathbf{B}(\mathbf{x}_t)\mathbf{x}_t \ , 
\quad
\mathbf{y}_t = \mathbf{C}(\mathbf{x}_t)\mathbf{h}_t \ .
\end{equation}
This allows Mamba to dynamically control how information is propagated and updated at each time step, improving its ability to model non-stationary sequences.
In addition, Mamba incorporates a continuous-time inductive bias via learned step sizes $\Delta t$, enabling adaptive temporal scaling of the state updates. 
This is particularly well-suited for Doppler-induced channel variations, which evolve continuously over time.

However, Mamba propagates information recursively through the hidden state. 
As a result, information from distant tokens must be compressed into $\mathbf{h}_t$, which can lead to degradation over long horizons.
Consequently, Mamba does not explicitly model pairwise interactions across the entire sequence as attention-based models do.
This can be limiting for CSI prediction, particularly in \emph{a) high-mobility scenarios}, where rapid channel variations introduce non-local dependencies across time and frequency, and in \emph{b) FDD settings}, where predicting DL CSI from UL CSI requires modeling complex cross-frequency relationships spanning distant tokens.

\begin{figure}[t]
    \centering
    \includegraphics[width=1\linewidth]{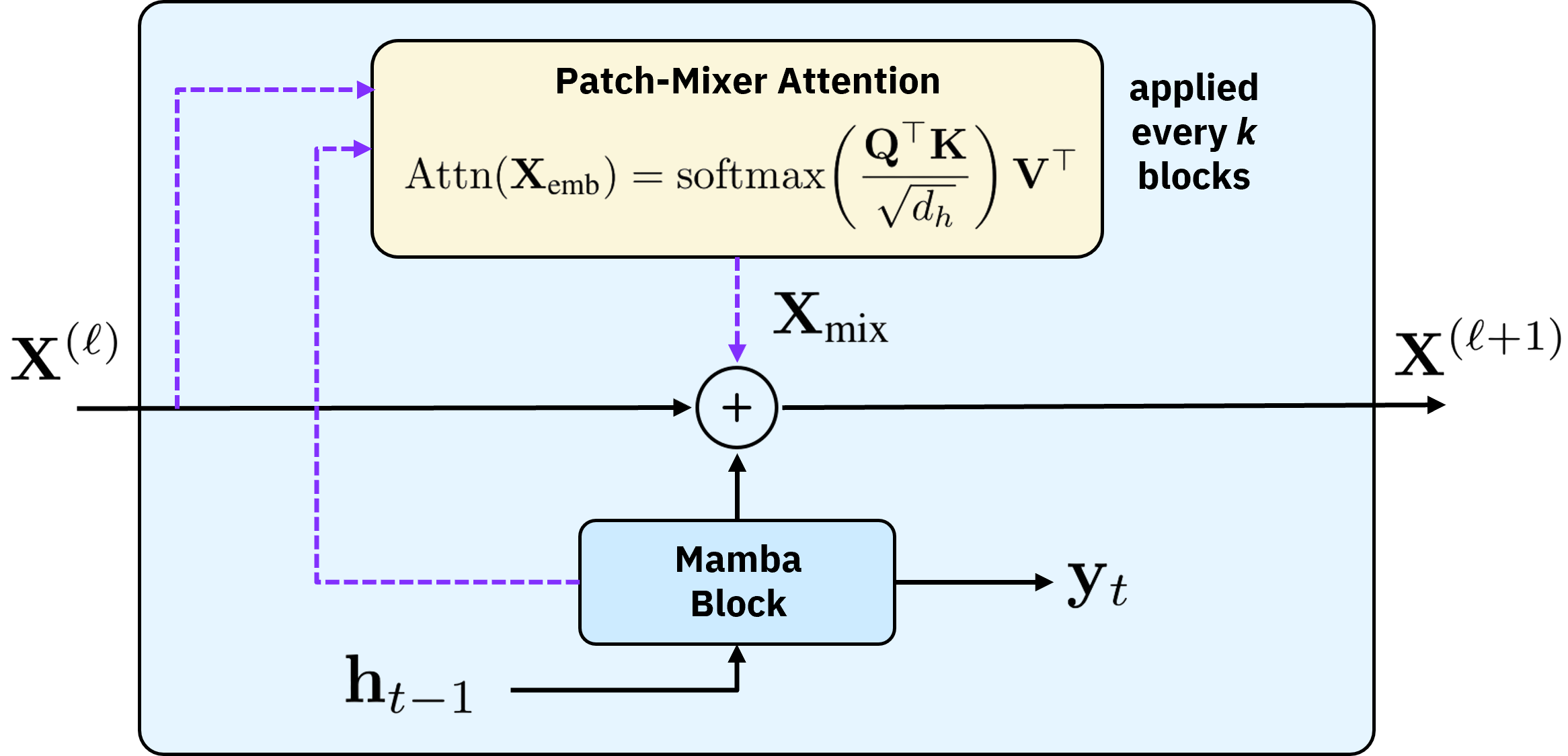}
    \caption{Hybrid-attention MambaCSP architecture at layer $l$.}
    \label{fig:MambaCSP}
    \vspace{-0.3cm}
\end{figure}

To address these limitations, we propose \textbf{MambaCSP}, a \emph{hybrid-attention} extension of the Mamba backbone (see \refig{fig:MambaCSP}).
Specifically, we introduce sparse \emph{patch-mixer attention layers}, which operate on the tokenized CSI patches every $k$ blocks and perform lightweight cross-token interactions using a small number of attention heads $H$ (typically 2--4).
Given $\mathbf{X}_{\text{emb}} \in \mathbb{R}^{F \times P'}$, patch-mixer attention is computed as
\begin{equation}
\mathrm{Attn}(\mathbf{X}_{\text{emb}})
=
\mathrm{softmax}\!\left(
\frac{\mathbf{Q}^\top \mathbf{K}}{\sqrt{d_h}}
\right)\mathbf{V}^\top \ ,
\end{equation}
where $d_h$ denotes the head dimension and $\mathbf{Q}$, $\mathbf{K}$, and $\mathbf{V}$ are the query, key, and value projections obtained from $\mathbf{X}_{\text{emb}}$ via learned linear mappings.
The resulting representations are concatenated and projected back via $\mathbf{W}_O \in \mathbb{R}^{F \times F}$, i.e.,
\begin{equation}
\mathbf{X}_{\text{mix}}
=
\mathrm{Concat}\!\bigl(
\mathrm{Attn}_1,\ldots,\mathrm{Attn}_H
\bigr)\mathbf{W}_O \ .
\end{equation}
The attention output is then inserted sparsely every $k$ Mamba blocks through a residual update, i.e.,
\begin{equation}
\mathbf{X}^{(\ell+1)}
=
\mathbf{X}^{(\ell)}
+
\mathbf{X}_{\text{mix}}^{(\ell)} \ ,
\qquad \ell \in \{k,2k,\dots\} \ .
\end{equation}
This restores global context modeling while maintaining near-linear complexity.
During training, patch-mixer attentions are learned jointly with projection and LayerNorm layers, while the majority of the Mamba 
backbone remains frozen.

\subsection{Computational Complexity and Efficiency}

Plain Mamba’s state-space updates process sequences one step at a time with recurrence.
Thus, for a token sequence of length $P'$ and hidden dimension $F$, compute scales linearly as $\mathcal{O}(P'F)$.
Moreover, since Mamba does not construct a full pairwise attention matrix and does not require KV caching, its memory usage also scales linearly with sequence length.

For our hybrid MambaCSP, patch-mixer attentions are inserted only intermittently and use a small number of heads.
Thus, for $L_M$ Mamba blocks and $\lfloor L_M/k \rfloor$ patch-mixer layers, where in practice $\lfloor L_M/k \rfloor \ll L_M$, the overall complexity is
\begin{equation}
\mathcal{O}(L_M P'F) + \mathcal{O}\!\left(\Bigl\lfloor \tfrac{L_M}{k} \Bigr\rfloor P'^2 F\right)
\;\approx\;
\mathcal{O}(L_M P'F) \ ,
\end{equation}
such that MambaCSP retains near-linear scaling overall.

In contrast, LLM backbones require self-attention in \emph{every layer}.
As a result, the forward computational cost scales quadratically as $\mathcal{O}(P'^2F)$, while memory is dominated by the attention matrix of size $P' \times P'$, i.e., $\mathcal{O}(P'^2)$.
Although freezing transformer layers reduces the number of trainable parameters during training, it does not remove the quadratic self-attention computation in the forward pass.
Therefore, MambaCSP provides a more favorable complexity--efficiency trade-off for long CSI histories and low-latency deployment.

\section{Experimental Setup}
\label{sec:experimental_setup}

To evaluate MambaCSP, we generate a comprehensive CSI dataset with QuaDRiGa~\cite{jaeckel2014quadriga} for the 3GPP TR~38.901 UMa NLOS channel.
The OFDM numerology and all remaining parameters follow our system model in \refsec{sec:system_model}.
The BS is placed at $(0,0,30)$\,m and is equipped with a $4\times4$ dual-polarized UPA.
For each UE, we simulate a frame of $N_{\text{slot}}=20$ slots, where the first $P=16$ slots form the history window and the remaining $L=4$ slots form the prediction window.
For each speed realization, we randomly place $10$ UEs in an annulus between 20\,m and 50\,m around a cluster center at $(200,0,1.5)$\,m and generate linear user tracks.
For DMRS, we employ a frequency-selective pilot pattern with temporal spacing $\Delta t_{\text{DMRS}}=2\Delta t$ and frequency spacing $\Delta k=1$ RB.
In total, we generate 8000 training, 2000 validation, and 10000 test samples for both TDD and FDD\footnote{Training code and data at: \url{https://github.com/aladinD/MambaCSP}}.

We benchmark our hybrid MambaCSP against plain Mamba, LLM-based prediction~\cite{liu2024llm4cp}, and classical neural baselines including CNN~\cite{CNN}, RNN~\cite{RNN}, and LSTM~\cite{LSTM}.
To ensure fairness, we employ a 130M parameter size Mamba 2 model~\cite{mambaHF}, whose scale is comparable to the 125M parameter size GPT-2~\cite{radford2019language} model used in~\cite{liu2024llm4cp}.
We further note that increasing the backbone model size or changing the architecture, e.g., to GPT-OSS~\cite{openai2025gptoss120bgptoss20bmodel} or other larger models, yields at most marginal gains for CSP, since the task does not primarily rely on semantic world knowledge or reasoning abilities (see Table~\ref{tab:backbone_scaling}).
We train each model for 10 epochs using Adam with $\beta=(0.9,0.999)$, batch size 256, and learning rate $10^{-3}$.
\section{Results and Discussions}
\label{sec:results}

We evaluate MambaCSP in terms of prediction accuracy and hardware efficiency, and perform additional ablations.

\subsection{Prediction Accuracy}

\begin{figure}[htbp]
    \centering
    \includegraphics[width=0.97\linewidth]{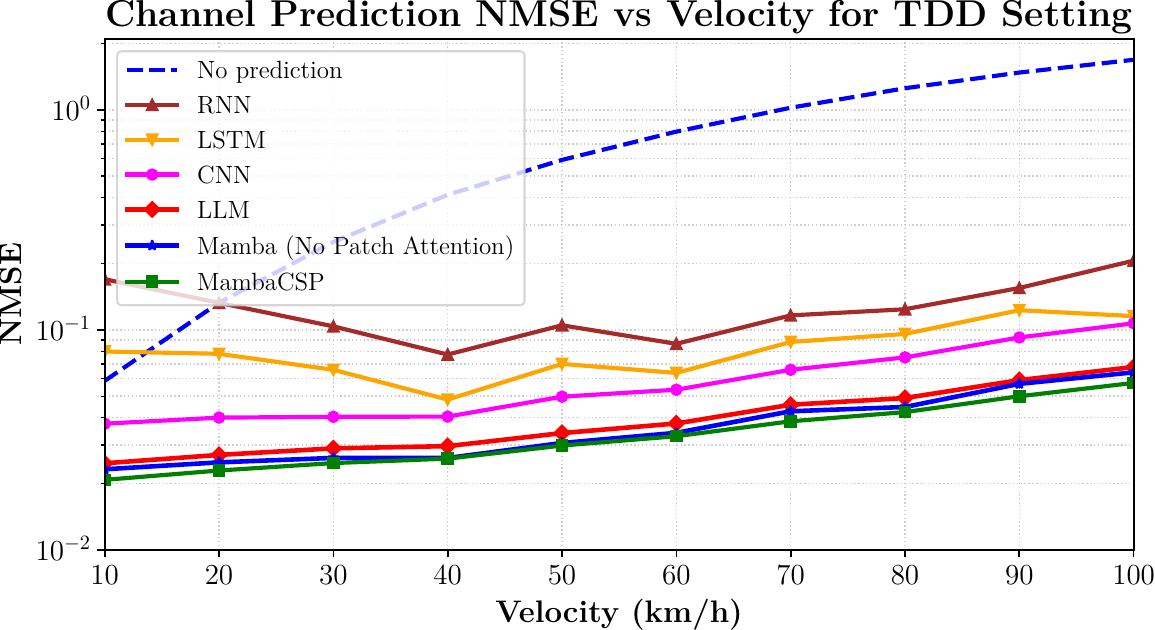}
    \caption{NMSE results for TDD. Across all approaches, the NMSE increases with increasing velocity. MambaCSP outperforms plain Mamba and the LLM.}
    \label{fig:nmse_tdd}
\end{figure}

\begin{figure}[htbp]
    \centering
    \includegraphics[width=0.97\linewidth]{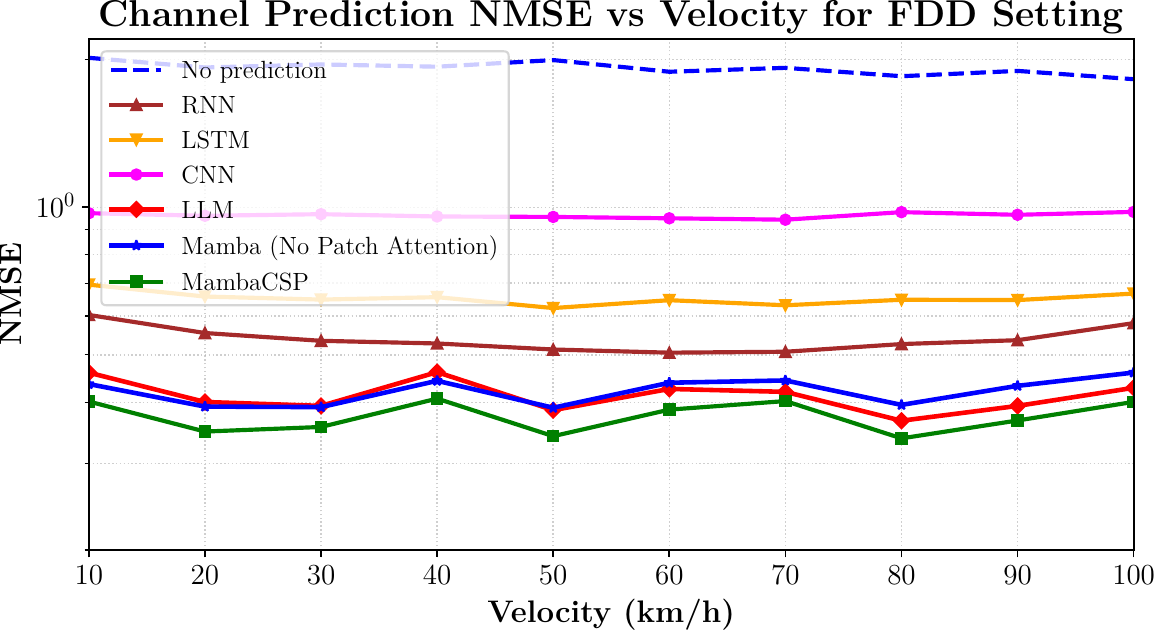}
    \caption{NMSE results for FDD. Across all approaches, the NMSE remains approximately flat. MambaCSP outperforms plain Mamba and the LLM.}
    \label{fig:nmse_fdd}
\end{figure}

\refig{fig:nmse_tdd} and \refig{fig:nmse_fdd} compare the NMSE of all considered baselines for the TDD and FDD settings, respectively, showing that MambaCSP outperforms LLM-based CSP in both duplex modes. 
For TDD, MambaCSP achieves 5\% lower NMSE than plain Mamba and 9\% lower NMSE than the LLM, indicating that state-space modeling already captures most relevant temporal structures in TDD, while patch-mixer attention provides a moderate but non-negligible gain.
For FDD, the relative benefit of MambaCSP becomes more pronounced.
It consistently outperforms plain Mamba by 17\% and the LLM by 12\% across the full velocity range, showing that patch-mixer attention is particularly beneficial in FDD when non-local cross-frequency dependencies arise due to the UL-to-DL prediction gap, which causes plain Mamba to underperform, particularly at higher velocities.
Furthermore, MambaCSP, plain Mamba, and the LLM significantly outperform CNN, RNN, and LSTM baselines, suggesting that sequence models are better at predicting spatio-temporal CSI dependencies.

\subsection{Throughput}

\begin{figure}[htbp]
    \centering
    \includegraphics[width=0.97\linewidth]{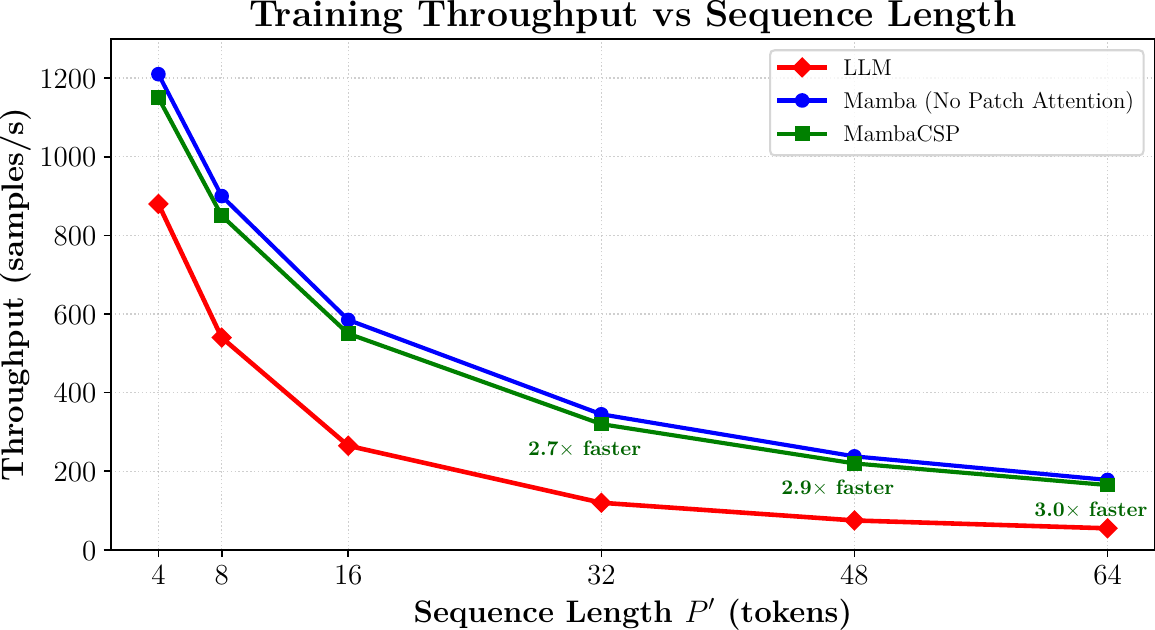}
    \caption{Throughput for different sequence lengths. MambaCSP achieves up to $3 \times$ more samples/s for longer CSI patches, comparable to plain Mamba. }
    \label{fig:throughput}
\end{figure}

\refig{fig:throughput} shows the training throughput as the sequence length $P'$ increases.
As expected, throughput decreases noticeably as the input history becomes longer.
Overall, LLM-based CSP degrades much faster than Mamba-based variants.
Specifically, MambaCSP achieves $2.7\times$, $2.9\times$, and $3.0\times$ higher throughput than the LLM at $P' \in \{32,48,64\}$, respectively, which is similar to plain Mamba. 
This shows that the proposed patch-level attention introduces only a minor throughput overhead.

\subsection{Memory Consumption}

\begin{figure}[htbp]
    \centering
    \includegraphics[width=0.97\linewidth]{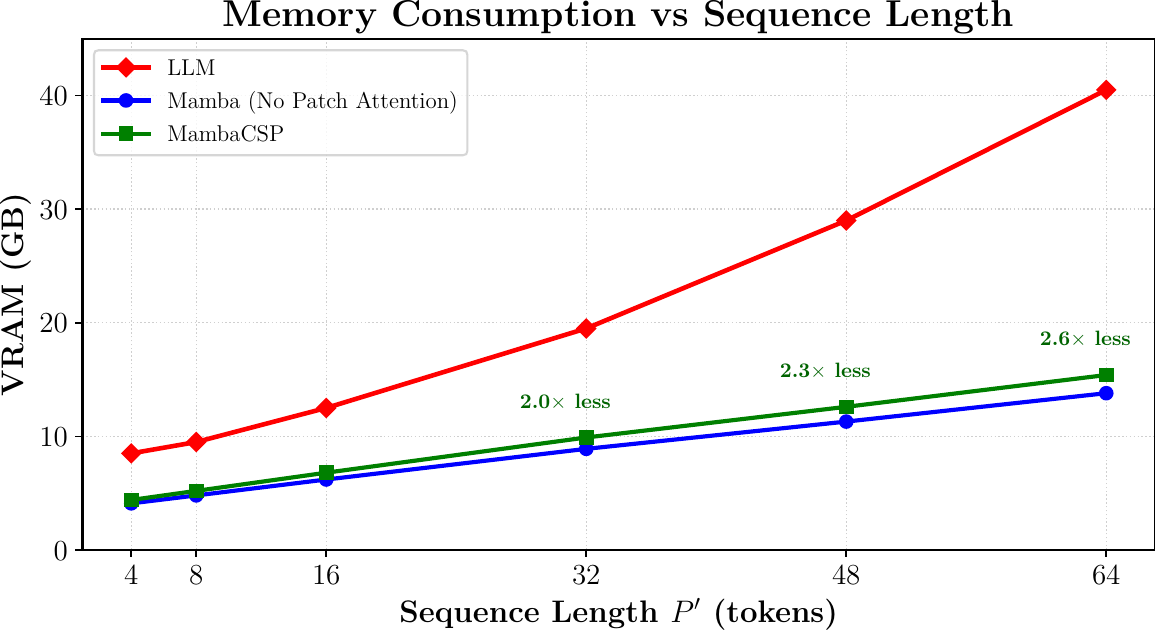}
    \caption{Memory consumption for different sequence lengths. LLM grows faster with sequence length due to KV caching. Mamba backbones remain linear.}
    \label{fig:memory}
\end{figure}

\refig{fig:memory} reports the peak memory usage when increasing the sequence length $P'$.
In line with our theoretical observations, the LLM exhibits a much steeper increase in memory consumption, driven by the quadratic attention matrix and KV caching.
For short input histories up to $P'=8$, the memory gap between LLM and Mamba-based models remains approximately constant, but it widens steadily as $P'$ grows.
Specifically, the LLM requires $2.0\times$, $2.3\times$, and $2.6\times$ more VRAM than the Mamba variants for $P' \in \{32,48,64\}$.
In contrast, plain Mamba and MambaCSP remain close and scale linearly in memory, substantiating our analysis that patch-level attention incurs only a negligible memory overhead.

\subsection{Inference Latency}

\begin{figure}[t]
    \centering
    \includegraphics[width=1\linewidth]{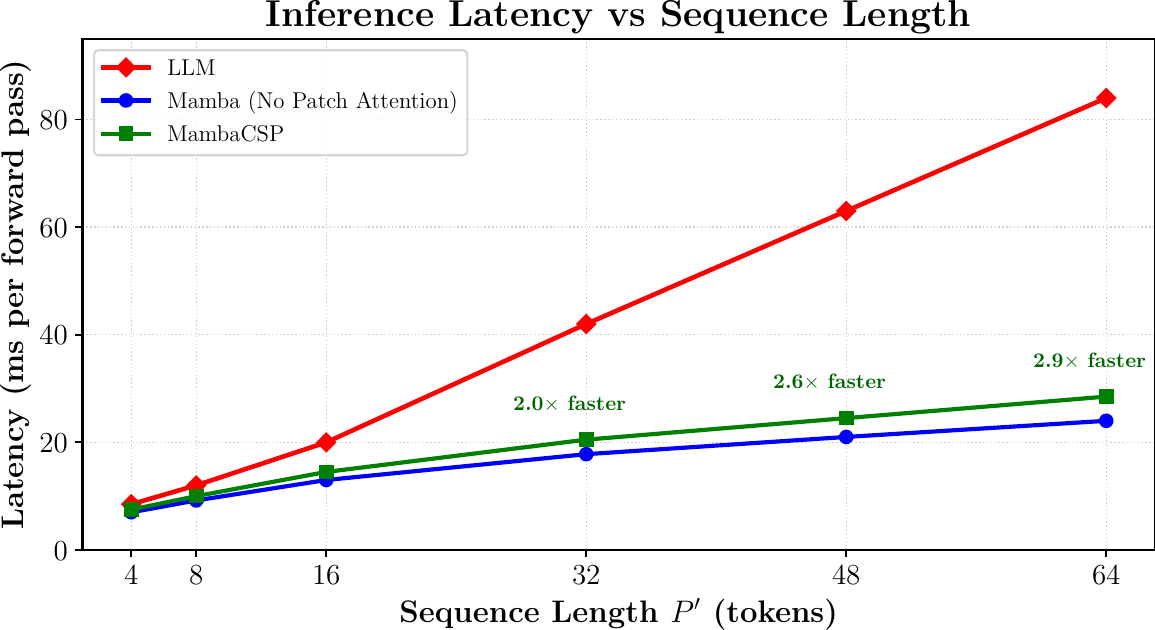}
    \caption{Latency for different sequence lengths. Mamba backbones scale linearly with longer sequence lengths compared to LLM (up to $3 \times$ slower).}
    \label{fig:latency}
    \vspace{-0.1cm}
\end{figure}

\refig{fig:latency} compares the inference latency per forward pass across sequence lengths.
Similar to memory usage, latency increases with $P'$ for all models.
MambaCSP is about $2.0\times$,

\newpage

\noindent
$2.6\times$, and $2.9\times$ faster than the LLM at $P' \in \{32,48,64\}$, respectively, while remaining close to plain Mamba, thereby preserving the latency benefits of the state-space backbone.

\subsection{NMSE vs Input Sequence Length}

Table~\ref{tab:nmse_vs_seqlen} reports the FDD NMSE for MambaCSP and LLM backbones as the input sequence length $P'$ increases.
Overall, a longer CSI history improves prediction (with diminishing returns at $P'=32$) as additional past observations expose more spatio-temporal structure that is useful for inferring future DL CSI.
This supports our motivation for increasing $P'$, especially in FDD, which benefits from longer-range CSI dependencies.
Similar trends can be observed for TDD.

\subsection{Ablations on MambaCSP Patch-Mixer Attention Layers}

Table~\ref{tab:patch_mixer_ablation} ablates the impact of the patch-mixer attention layers in MambaCSP.
Overall, injecting cross-attention with $H=2$ heads every $k=4$ Mamba blocks is sufficient to restore non-local sequence modeling capabilities for complex CSI patches, while providing a favorable trade-off between prediction accuracy and efficiency.
Further increasing the number of heads or inserting attention more frequently yields only marginal additional gains at the cost of increased complexity.

\ 

In summary, our results demonstrate that MambaCSP achieves a substantially better accuracy--efficiency trade-off than LLM-based CSP, particularly for longer CSI histories. 
Its lower latency and memory footprint make it especially attractive for real-time deployment on low-resource devices.
Further reducing memory and compute through quantization and model pruning is an interesting direction for future work.
\section{Conclusion}
\label{sec:conclusion}

In this paper, we proposed MambaCSP, a hybrid-attention state space model architecture as a hardware-efficient alternative to LLM-based CSI prediction.
MambaCSP replaces the LLM backbone in existing CSP pipelines with a Mamba-based sequence model and augments it with sparse patch-mixer attention layers to better capture long-range CSI token dependencies while preserving the linear-time efficiency of state space models.
Our experiments show that MambaCSP achieves better accuracy than LLM-based prediction in both

\newpage

\begin{table}[htbp]
\centering
\caption{NMSE versus sequence length $P'$ for LLM and MambaCSP (FDD). Increasing $P'$ improves prediction with diminishing returns at larger contexts.}
\label{tab:nmse_vs_seqlen}
\resizebox{0.95\columnwidth}{!}{%
\begin{tabular}{c|cc|c}
\hline
\textbf{Seq. Length $P'$} & \textbf{LLM NMSE} & \textbf{MambaCSP NMSE} & \textbf{Relative Gain} \\
\hline
4  & 0.503 & 0.468 & 7.0\% \\
8  & 0.462 & 0.422 & 8.7\% \\
16 & 0.425 & 0.366 & 13.9\% \\
32 & 0.392 & 0.331 & 15.6\% \\
48 & 0.381 & 0.318 & 16.5\% \\
64 & 0.374 & 0.309 & 17.4\% \\
\hline
\end{tabular}%
}
\end{table}

\vspace{-0.4cm}

\begin{table}[htbp]
\centering
\caption{Ablation of the patch-mixer attention layers in MambaCSP. Sparse attention improves FDD more than TDD. In practice, we use $k=4$ and $H=2$ as a favorable accuracy-efficiency trade-off.}
\label{tab:patch_mixer_ablation}
\resizebox{0.95\columnwidth}{!}{%
\begin{tabular}{c c c c c}
\hline
\textbf{Patch-Mixer} & \textbf{Interval $k$} & \textbf{Heads $H$} & \textbf{TDD NMSE} $\downarrow$ & \textbf{FDD NMSE} $\downarrow$ \\
\hline
None & -- & -- & 0.0374 & 0.4222 \\
Yes  & 4  & 2  & 0.0346 & 0.3750 \\
Yes  & 4  & 4  & \textbf{0.0345} & \textbf{0.3736} \\
Yes  & 2  & 4  & 0.0345 & 0.3731 \\
\hline
\end{tabular}%
}
\end{table}

\vspace{-0.4cm}

\begin{table}[htbp]
\centering
\caption{Comparison of different backbones. Larger models provide only marginal gains for CSP despite substantially higher memory requirements.}
\label{tab:backbone_scaling}
\resizebox{0.9\columnwidth}{!}{%
\begin{tabular}{lcccc}
\hline
\textbf{Backbone} & \textbf{Params} & \textbf{VRAM (GB)} & \textbf{TDD NMSE} $\downarrow$ & \textbf{FDD NMSE} $\downarrow$ \\
\hline
GPT-2 Small   & 124M & 6.8  & 0.050 & 0.425 \\
GPT-OSS       & 1.3B & 18.9 & 0.049 & 0.423 \\
Mamba2-130M   & 130M & 5.1  & 0.049 & 0.422 \\
Mamba2-780M   & 780M & 11.8 & 0.048 & 0.420 \\
\hline
\end{tabular}%
}
\end{table}

\noindent
TDD and FDD settings, while providing substantially higher throughput, lower memory consumption, and lower inference latency.
Our findings thus point to a key takeaway:
LLMs are not the final architectural choice for long-context CSI prediction, and combining efficient state space models with lightweight attention is a promising path toward scalable and real-time AI-native CSI prediction in future wireless systems.


\bibliographystyle{IEEEtran}
\bibliography{IEEEabrv,bibliography}

\end{document}